\documentclass{nature_bf2}
\usepackage{graphicx}
\usepackage[usenames]{color}

\title{Fractal multi-level organisation of human groups in a virtual world} 
\author{Benedikt Fuchs$^1$, Didier Sornette$^{2}$ \& Stefan Thurner$^{1,3,4}$}

\begin{document}

\maketitle

\begin{affiliations}
 \item Section for Science of Complex Systems, Medical University of Vienna, Spitalgasse 23, A-1090, Austria
 \item Department of Management, Technology and Economics, ETH Z\"urich, Scheuchzerstrasse 7, CH-8092 Zurich, Switzerland
 \item Santa Fe Institute, 1399 Hyde Park Road, Santa Fe, NM 87501, USA
 \item International Institute for Applied Systems Analysis, Schlossplatz 1, A-2361 Laxenburg, Austria
\end{affiliations}

\begin{abstract} 
Humans are fundamentally social. They have progressively dominated 
their environment by the strength and creativity provided by and within their grouping.
It is well recognised that human groups are highly structured,
and the anthropological literature has loosely classified them according
to their size and function, such as
support cliques, sympathy groups, bands, cognitive groups, tribes, linguistic groups and so on\cite{Dunbar1995,Hill2003,Dunbar1993}.
Recently, combining data on human grouping patterns in a
comprehensive and systematic study, Zhou et al.\cite{Zhou2005}
identified a quantitative discrete hierarchy of group sizes with a preferred scaling 
ratio close to $3$, which was later confirmed for hunter-gatherer groups\cite{Hamilton07}
and for other mammalian societies\cite{Hill2008}.
Using high precision large scale Internet-based social network data, 
we extend these early findings
on a very large data set. We analyse the organisational structure of a complete,
multi-relational, large social multiplex network of a human society consisting of about 400,000 
odd players of a massive multiplayer online game for which we know 
all about the group memberships of every player. Remarkably, the online players
exhibit the same type of structured hierarchical layers as
the societies studied by anthropologists, where each of these layers is three to four times 
the size of the lower layer, as illustrated in Fig. 1. Our findings suggest that the hierarchical organisation of human society
is deeply nested in human psychology.
\end{abstract}

We analyse comprehensive data from a society, consisting of the players of the massive multiplayer online game (MMOG) 
\emph{Pardus}\footnote{www.pardus.at}. Such online platforms provide a new way of observing hundreds of thousands 
of interacting individuals who are engaged in social and economic activities,
enabling quantitative socioeconomic research\cite{Bainbridge2007,Castronova2005,Szell2010msd,Klimek2013,Szell2010mol,Thurner2012,Szell2013,Szell2012}. 
Complementing traditional methods of social science such as small-scale questionnaire-based approaches, MMOGs allow 
the study of complete societies, which are free of any interviewer bias or laboratory effects since users are not aware 
that their actions are logged during playing. Moreover, we have complete information about all players at any point in time. 

Extensive studies on Pardus have shown remarkable similarities between this virtual world and real-world societies, 
in terms of network structure\cite{Szell2010msd,Klimek2013,Szell2010mol}, social behaviour\cite{Thurner2012,Szell2013}, and mobility patterns\cite{Szell2012}. 
Players in Pardus control characters (avatars) who `live' in a virtual, futuristic universe and interact with others in a 
multitude of ways to achieve their self-posed goals. Since the game went online in 2004, more than 400,000 people 
have played it. 
Pardus provides an internal one-to-one messaging system comparable to emails and players can express their sympathy 
toward other players by marking them as friend. 

Players interact with each other in a multitude of ways, creating a social multiplex network\cite{Szell2010mol}.
Next to small friendship and support groups, they can explicitly form social groups and register these as \emph{alliances}, 
which grants tools to facilitate administration of the groups. Averaged over our whole data set (see Methods), alliances have an average size of 24.7. 
There is no upper bound for alliances sizes, however we find that the largest alliance has 136 members, 
which is remarkably close to Dunbar's number\cite{Dunbar1993}.
The largest type of registered groups in the game are `political' \emph{factions}, with about 2,000 members. 
The number of factions in the game is limited to three, their relative sizes and numbers of memberships  are not fixed.
Factions claim territories and can wage war against other factions. Each alliance may decide to belong to 
one of the three factions. Players may decide against membership in any of theses groups.
The average size  at any time of the entire society is about $N=7,065$ active players.

The possibility for diverse levels of organisation gives rise to a hierarchical society with a complex structure, which 
we quantify in two complementary ways, first using the Horton-Strahler measure of branching complexity,
and second by studying directly the structure of the distribution of group sizes.

We construct the Horton-Strahler orders for all players (see Methods). 
Fig. 1 shows a part of the social network on Pardus where the different Horton orders are defined.
The  innermost layer,  Horton order $h=1$, is the trivial group consisting of one person, the `ego'.
Layer 2 ($h=2$) contains closest friends of the ego, defined by both a friendship marking and at least one communication event within the last 30 days.
Layer 3 ($h=3$) includes more casual relations, in particular all players that ego has marked as a friend, or by whom ego was marked as friend.
Layer 4 ($h=4$) contains the fellow alliance members of the ego. 
Layer 5 ($h=5$), corresponding to the communication clusters, 
is obtained by applying a community detection algorithm (Louvain algorithm)\cite{Blondel2008,Rubinov2010} to the communication network of the players
(see Methods).
We tested explicitly that layer 5 is an organisational layer in its own right, 
whose communities are predominantly subsets of the factions ($h=6$) and supersets of the alliances ($h=4$, see Methods).
Layer 6 ($h=6$) contains the three factions, and layer 7  ($h=7$) is the entire society.

\begin{figure}[htb]
 \centering
	 \includegraphics[width=8.9cm]{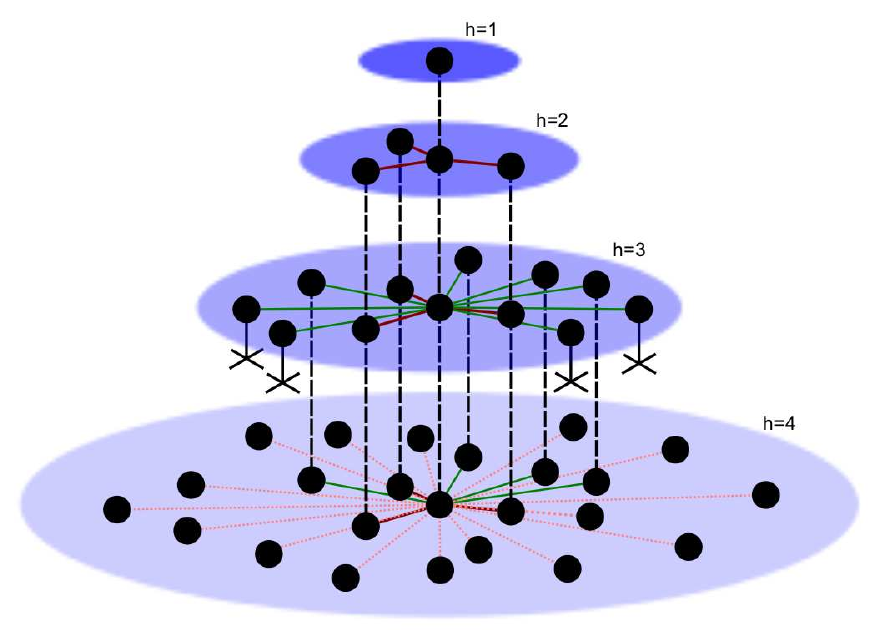}
 \caption{Ego-network of one particular player showing hierarchical organisation.
Blue ellipses depict the various layers of organisation. Dots represent players; dashed lines 
connect identical players across layers; crosses denote players that are not present in the next layer. Thick dark red lines represent strongest ties, forming $G_i^2$, 
green lines represent friendship links, forming $G_i^3$, and dotted pink lines mark membership in a common alliance.
The layers contain 1, 4, 12, and 24 individuals, respectively. Layer 3 (friends) is typically not a subset of layer 4 (alliance).
For clarity, only links to the ego are drawn.}
 \label{fig:hierarchicalNW}
\end{figure}

\begin{figure}[tbp]
 \centering
 \includegraphics[width=12cm]{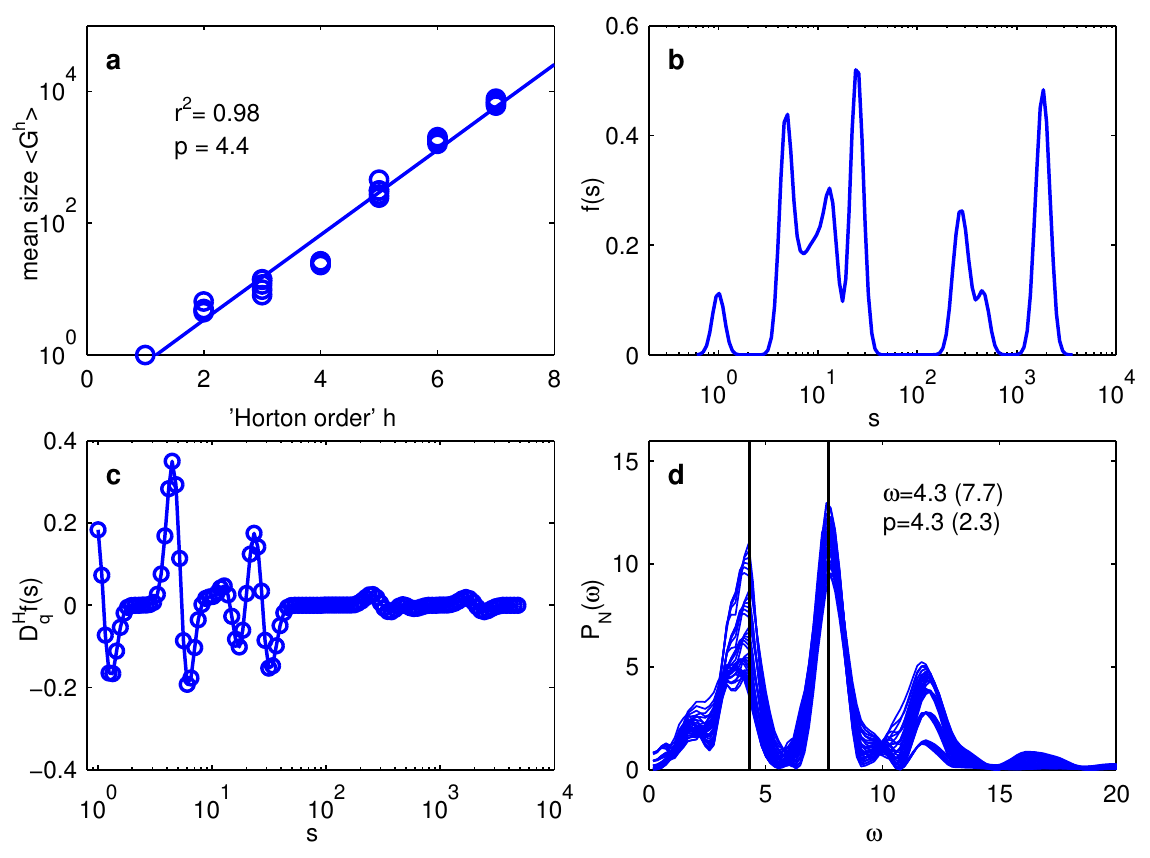}
 \caption{Analysis of group size scaling:
 a Horton plot: average size of groups per order.
 b Estimated probability density of group sizes in Pardus, obtained as Gaussian kernel estimation with bandwidth $\sigma=0.14$ acting on $\ln(s)$ (see Methods).
 c Generalized $(H,q)$ derivative of $f(s)$ for $H=0.5$ and $q=0.8$ (see Methods). 
 d Lomb periodogram of the $(H,q)$-derivative of $f(s)$ for different values of $H$ and $q$ (see Methods). %
Peaks at $\omega\approx 4.3  \; (7.7)$ (marked with black vertical lines) correspond to scaling ratios $p = \exp(2\pi/\omega) \approx 4.3 \; (2.3)$.
 }
 \label{fig:analysis}
\end{figure}

Following Hill et al.\cite{Hill2008}, we calculate the average group size at Horton order $h$, $\langle G^h \rangle$. 
Figure \ref{fig:analysis}~a shows that $\langle G^h \rangle \sim p^{h}$, with a scale ratio of $p = 4.4$.

A second independent way to affirm discrete scale invariant structure is obtained by directly analysing the distribution of group sizes, 
following the approach presented by Zhou et al.\cite{Zhou2005}. Using a Gaussian kernel estimator of the probability 
density $f(s)$ (shown in Fig. \ref{fig:analysis}~b) of player group sizes in our data, we calculate the generalized $(H,q)$-derivative\cite{ZhouSor02,ZhouSor03} of $f(s)$,
 which generalises the $q$-derivative\cite{Erzan1997,Erzan1997b}, for multiple values of 
$H$ and $q$, see Fig. \ref{fig:analysis}~c. The parameter $H$ stands for the Hurst exponent used to rescale the derivative, while $q$ controls the 
scale factor of the $q$-derivative. Coupled with the Lomb-periodogram\cite{Press2007}, the $(H,q)$-derivative
has been shown to be very efficient for identifying log-periodicity in signals\cite{ZhouSor02,ZhouSor03}, which is the observable
signature of discrete scale invariance\cite{SornetteDSI98}. As shown in Fig. \ref{fig:analysis}~d, the Lomb periodogram of 
the $(H,q)$-derivative of $f(s)$ gives a highly significant peak\cite{ZhouSornetteSig02} at the angular log-frequencies 
$\omega = 4.3$, corresponding to a scaling ratio $p = \exp(2\pi/\omega) =  4.3$.
One can further clearly see the second and third harmonics, which gives additional support for the existence of log-periodicity\cite{zhousorturb02},
and therefore hierarchical, and discrete scale invariance.

\begin{methods}

\subsection{Data.}\label{sec:data}
Pardus is partitioned into three independent games, called 'universes'. Here, we focus on one of them, the 'Artemis' universe.
In the game, we have complete information on a multitude of temporal social networks, including the friendship-, communication-, and trading networks\cite{Szell2010mol}. 
Data are available over 1238 days. We take snapshots of the friendship- and communication network and of group affiliations on days 240, 480, 720, 960, and 1200, respectively.
In more formal terms, we have a multiplex $\mathcal{M}_{ij}^\alpha(t)$, where $\alpha$ indicates the type of the link, here friendship and communication.
$\mathcal{M}_{ij}^\mathrm{friend}(t)=1$ if $i$ has marked $j$ as friend before $t$ (and has not revoked this marking since) and zero otherwise.
$\mathcal{M}_{ij}^\mathrm{comm.}(t)=1$ if $i$ has sent a message to $j$ in the time $[t-30\mathrm{d}, t]$ and zero otherwise.
Further, we consider the symmetrisation of the multiplex: $\hat{\mathcal{M}}_{ij}^\alpha = 1$ if $\mathcal{M}_{ij}^\alpha = 1$ or $\mathcal{M}_{ji}^\alpha = 1$.
Groups are defined at 6 layers, starting with the ego ($h=1$), where $G_i^h$ is the group of layer $h$ to which $i$ belongs. 
Support cliques ($h=2$), are defined as the set of an individual's friends with whom he has communicated at least once within the last month: $G_i^2(t) = \lbrace j : \hat{\mathcal{M}}_{ij}^\mathrm{friend}(t)=\hat{\mathcal{M}}_{ij}^\mathrm{comm.}(t)=1 \rbrace$.
Sympathy groups ($h=3$) are defined as the set of an individual's friends:
$G_i^3(t) = \lbrace j : \hat{\mathcal{M}}_{ij}^\mathrm{friend}(t)=1 \rbrace$.
Layer 4 consists of the so-called 'alliances' $(h=4)$, which are clubs that can be created in the game and where all memberships are known. The same is true for the 'factions' $(h=6)$.
An additional layer of grouping $(h=5)$ is found by applying the Louvain algorithm\cite{Blondel2008,Rubinov2010} to 
$\hat{\mathcal{M}}_{ij}^\mathrm{comm.}(t)$. Note that the Louvain algorithm confirms the
other lower layers $h=1$ to $h=4$. The last layer is the whole society $(h=7)$.
We consider only alliances with at least three members.
 $G_i^h(t)$ might be empty for $h>1$.

\subsection{Layer 5: communication clusters.}\label{sec:comm}
Layer 5 is obtained by applying a community detection algorithm (Louvain algorithm\cite{Blondel2008,Rubinov2010}) to the 
communication network of the players. 
 We detect communities that contain 292 players each (ignoring communities with less than three members). 
Results for every day in our data set are obtained from averages over five runs of the algorithm.  
When comparing the layer 5 communities to the factions, we find that about 76\% of the members of a layer 5 community are in the same faction.
Comparing layer 5 communities to the alliances we find that about 84\% of the members of an alliance are in the same layer 5 community on average.
To further quantify the similarity between communities found by the Louvain algorithm and the factions and alliances, we calculate the 
Fowlkes-Mallows index\cite{Fowlkes1983,Halkidi2001} $\mathcal{F}$ (see below).
$\mathcal{F}$ compares two results of community labelling (clusterings, partitions).
For identical clusterings, $\mathcal{F}=1$,
for totally unrelated clusterings, $\mathcal{F}\to0$, given the number of clusters is large. 
Here we compare layer 5 with the factions (layer 6) and the alliances (layer 4).
As a null model we generate random communities of the same sizes as those found by the Louvain algorithm:
each community labelling (as found in any of the five runs of the Louvain algorithm) is reshuffled ten times, 
and the respective Fowlkes-Mallows indices for layer 5 -- factions and layer 5 -- alliances are computed.
$\mathcal{F}_\mathrm{shuffle}$ is defined as the average over the five iterations of the Louvain algorithm, the ten shuffled versions, and the five days of observation.
For layer 5 -- factions, we find $\mathcal{F}=0.50$, with $\mathcal{F}_\mathrm{shuffle}=0.21$,
which suggests that the detected communities are predominantly subsets of the factions. 
For the layer 5 -- alliances case we get
$\mathcal{F}=0.28$, and $\mathcal{F}_\mathrm{shuffle}=0.041$, implying that the layer 5 communities are also mainly supersets of the alliances.
These results indicate that layer 5 is indeed an organisational layer in its own right, located between the factions (layer 6) and the alliances (layer 4).

\subsection{Fowlkes-Mallows index $\mathcal{F}$.}
To validate the communities found by the Louvain algorithm, we compare them to the factions and alliances by 
means of the Fowlkes-Mallows index\cite{Fowlkes1983,Halkidi2001} $\mathcal{F}$, which is defined as:
\begin{displaymath}
 \mathcal{F} \equiv \frac {TP}{\sqrt{(TP+FP) (TP+FN)}} \quad ,
\end{displaymath}
where $TP$ is the number of pairs of elements that are in a common community in both compared clusterings,
$FP$ is the number of pairs that are in a common community in clustering 1, but belong to two different communities in clustering 2.
$FN$ is the number of pairs that are found in a common community in clustering 2, but belong to two different communities in clustering 1.

\subsection{Gaussian kernel estimator.} For a smooth estimation of $f(s)$ from our $N$ data points $s_i$, we use the Gaussian kernel estimator $f(\ln(s)) = \frac 1 N \sum_{i=1}^N\mathcal{N}(\ln(s)-\ln(s_i), \sigma)$, 
where $\mathcal{N}(0,\sigma)$ is a zero-mean Gaussian distribution with standard deviation $\sigma=0.14$.

\subsection{Generalized $(H,q)$-derivative.} is defined as\cite{Erzan1997,Erzan1997b} $D^H_qf(s) \equiv \frac{f(s)-f(qs)}{\left[(1-q)s\right]^H}$.
It allows for an adaptive de-trending and enhances possible discrete scale structures.

\subsection{Lomb periodogram.} For the frequency analysis, we use the Lomb periodogram\cite{Press2007},
 which provides an ideal spectral analysis for unevenly sampled data, as occurs when using
the logarithm of sizes. See Fig. 3 that illustrates the whole process of recovering the preferred scaling ratio
using the Lomb periodogram applied to the  $(H,q)$-derivative  of the kernel estimation of the density
distribution of a noisy log-periodic signal.

\begin{figure}[p]
 \centering
 \includegraphics[width=8.9cm]{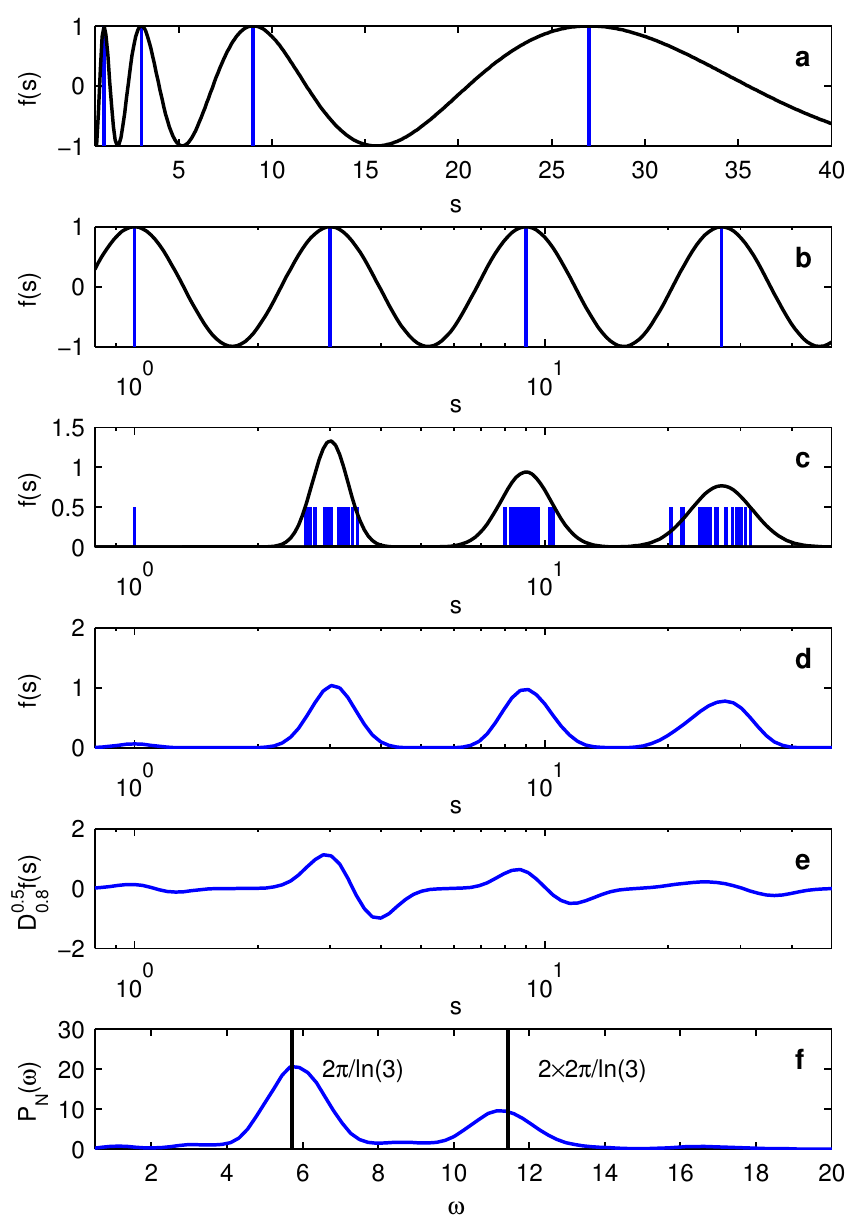}
 \caption{Detection of log-periodicity (illustration): %
a Data points with factor three between each other (blue), log-periodic function $\cos(\omega_0\ln(s))$ with $\omega_0 = 2\pi/\ln(3)\approx 5.72$ (black). %
b Same as (a), but with logarithmic x-axis to visualise the log-periodicity. %
c $\ln(s)$ is perturbed, i.e. drawn from a (sum of) normal distribution(s) with mean $\ln(3)$ and variance 0.1 (blue). %
 Black: Analytical probability density for the data. %
d Probability density as inferred from the data by Gaussian kernel estimation of $\ln(s)$ with bandwidth 0.1. %
e Generalized (H,q)-derivative of $f(s)$, with $q=0.8$ and $H=0.5$. %
f Lomb periodogram of $D^H_qf(s)$ as function of $\ln(f(s))$. The main peak is close to the expected value $\omega_0$ (marked in black). %
Additionally, a peak close to the second harmonic $2\omega_0$ is visible.}
 \label{fig:method}
\end{figure}

\end{methods}

\begin{addendum}
 \item ST and BF acknowledge support from the Austrian Science Fund FWF P23378.
 \item[Author Contributions] BF, DS, and ST devised the concept. BF analysed the data.
BF, DS, and ST interpreted the results and wrote the manuscript.
 \item[Author Information] The authors declare that they have no competing financial interests.
 Correspondence and requests for materials should be addressed to ST (stefan.thurner@meduniwien.ac.at).
\end{addendum}

\end{document}